\documentclass[11pt,twoside]{article}

\usepackage{asp2010}
\usepackage{graphicx}
\def\be{\begin{equation}}
\def\ee{\end{equation}}
\def\lsim{\mathrel{\hbox{\rlap{\hbox{\lower4pt\hbox{$\sim$}}}\hbox{$<$}}}}
\def\gsim{\mathrel{\hbox{\rlap{\hbox{\lower4pt\hbox{$\sim$}}}\hbox{$>$}}}}

\resetcounters 

\begin{document}

\resetcounters

\title{Cosmological Backgrounds of Gravitational Waves and eLISA}

\author{Jean-Fran\c{c}ois Dufaux 
\affil{APC, Univ. Paris Diderot, CNRS/IN2P3, CEA/Irfu, Obs. de Paris, Sorbonne Paris Cit\'e, France}}

\begin{abstract} 
We review cosmological backgrounds of gravitational waves with a particular attention to the scientific potential of the eLISA/NGO mission. After an overview of cosmological backgrounds and detectors, we consider different cosmological sources that could lead to an observable signal. We then study the backgrounds produced by first-order phase transitions and networks of cosmic strings, assessing the prospects for their detection.
\end{abstract}

\begin{center}
\emph{To appear in the proceedings of the 9th LISA Symposium}
\end{center}

\section{Introduction} 

Because of the weakness of the gravitational interaction, gravitational waves (GW) propagate freely during essentially 
all the cosmic history. Once produced in the early universe, they simply red-shift with the cosmological expansion, retaining their initial frequency dependence. This allows GW detectors to directly probe scales, epochs and physical phenomena that are not accessible in any other way~(Refs.~\cite{Allen:1996vm}, \cite{Maggiore:1999vm}, \cite{Buonanno:2003th}, \cite{Hogan:2006va}).

In this contribution, based on \cite{Binetruy:2012ze}, we review cosmological GW backgrounds with a particular attention to those backgrounds that could be observable by the eLISA mission~(Refs.~\cite{AmaroSeoane:2012km}, \cite{AmaroSeoane:2012je}). eLISA (also named NGO by ESA) is the new, European version of LISA that is now being proposed~\footnote{See http://www.elisa-ngo.org/ for more details.}. We start with an overview of cosmological backgrounds and detectors in Section~\ref{CosmoBack}. In Section~\ref{CosmoSou}, we discuss several cosmological sources of GW. We then study the backgrounds produced by first-order phase transitions (Section~\ref{CosmoPT}) and cosmic strings (Section~\ref{CosmoStrings}), assessing the prospects for their detection. We conclude in Section~\ref{CosmoConclu}. 

\emph{Notation:} we work in units where $c = k_B = 1$.

\section{Cosmological Backgrounds and Detectors}
\label{CosmoBack}

A cosmological source of GW typically consists of many uncorrelated and unresolved events, which produce a stochastic GW background. In general, such a background is expected to be nearly Gaussian, isotropic, stationary and 
unpolarized~(Refs.~\cite{Allen:1996vm}, \cite{Maggiore:1999vm}). Its main property is then its frequency spectrum. The quantity that is usually considered to characterize cosmological backgrounds is the spectrum of GW energy density per logarithmic frequency interval, normalized to the critical energy density today
\be
\label{CosmoOmega}
h^2\,\Omega_{gw}(f) = \frac{h^2}{\rho_c} \, \frac{d \rho_{gw}}{d \log f}
\ee 
where $f$ is the present-day GW frequency, $\rho_{gw}$ the present-day GW energy density, 
$\rho_c = 3 H_0^2 / (8 \pi G)$ is the critical energy density today, and $h \simeq 0.72$ parametrizes the small uncertainty in the value of the Hubble constant today ($H_0 = 100\,h\,$km/s/Mpc).

There are currently a number of upper bounds on cosmological GW backgrounds, in very different frequency ranges. The amount of GW that is present at the time of Big Bang Nucleosynthesis (BBN) is constrained not to spoil the successful predictions for the light element abundances~(Ref.~\cite{Maggiore:1999vm}). Although this bound applies to the total GW energy density 
$\rho_{gw}$, neglecting the unlikely possibility of a GW spectrum with a very narrow peak one can take as a rule of thumb $h^2 \Omega_{gw} < 10^{-5}$ for $f \gsim 10^{-10}$ Hz ($f \sim 10^{-10}$ Hz corresponds to the comoving Hubble scale at the time of BBN). Similarly, the total GW energy density that is present at the time of decoupling of the Cosmic Microwave Background (CMB) is constrained by precise measurements of the CMB power spectrum~(Ref.~\cite{Smith:2006nka}). The current bound is also of the order of $h^2 \Omega_{gw} < 10^{-5}$, but it covers a larger frequency range, $f \gsim 10^{-15}$ Hz. At the lowest observable frequencies, $f \sim 10^{-18} - 10^{-16}$ Hz, cosmological GW are constrained by their contribution to the large-angle fluctuations of the CMB temperature~(Ref.~\cite{Maggiore:1999vm}). The strongest bound is obtained at $f \sim 10^{-16}$ Hz: $h^2 \Omega_{gw} < \mathrm{few} \times 10^{-14}$. The sensitivity to cosmological GW in this frequency range may improve significantly in the future with measurements of the CMB polarization (Refs.~\cite{Kamionkowski:1996zd}, \cite{Seljak:1996gy}). GW backgrounds are also constrained by the very accurate timing of msec pulsar arrays: Ref.~\cite{vanHaasteren:2011ni} recently obtained the upper bound $h^2 \Omega_{gw} \lsim 6 \times 10^{-9}$ at $f = 4 \times 10^{-9}$ Hz (up to a weak dependence on the slope of the spectrum). The sensitivity of pulsar timing arrays to stochastic GW backgrounds is expected to improve significantly in the future~(Ref.~\cite{Jenet:2006sv}). Finally, the S5 run of LIGO allowed to put the upper bound~(Ref.~\cite{Abbott:2009ws}) $h^2 \Omega_{gw} \lsim 3.6 \times 10^{-6}$ at $f \approx 100$ Hz, which improves upon the BBN and CMB bounds in this frequency range. Again, the sensitivity to stochastic GW backgrounds in this frequency range will improve significantly in the near future with the advanced ground-based interferometers.  

For ground-based interferometers, the best way to disentangle a stochastic background from the instrumental noise is to cross-correlate the output of different detectors (see \cite{Allen:1996vm}, \cite{Maggiore:1999vm} and references therein). This is not possible for eLISA since there is a single instrument. In this case, the spectral density of the signal must be larger than the one of the noise. However, the fact that eLISA is designed to probe lower frequencies ($f \sim 10^{-4} - 10^{-1}$ Hz) is an important advantage, because the spectral density of the signal $S_h(f)$ is related to $\Omega_{gw}$ as 
$\Omega_{gw} \propto f^3\,S_h(f) / H_0^2$ (Ref.~\cite{Maggiore:1999vm}). Thus, for a given sensitivity to $S_h(f)$, the sensitivity to the GW energy density quickly improves as the frequency decreases. Compared to the original LISA mission, it is more difficult for eLISA to disentangle a stochastic background from the instrumental noise, all the more because the eLISA Mother-Daughter configuration, providing only two measurement arms, does not allow to use Sagnac calibration~(Refs.~\cite{Tinto:2001ii}, \cite{Hogan:2001jn}). Luckily as we will see, for sources like phase transitions and cosmic strings, the spectral dependence of the signal is well predicted and may allow to distinguish cosmological backgrounds as long as they lie above the sensitivity curve. In Sections~\ref{CosmoPT} and \ref{CosmoStrings}, we consider the sensitivity curve of eLISA computed from the expected instrumental noise and after removing the confusion noise generated by unresolved galactic binaries (Ref.~\cite{AmaroSeoane:2012km}).

\section{Cosmological Sources of GW}
\label{CosmoSou}

A cosmological source emits GW with a characteristic wavelength $\lambda_*$ that is smaller than the Hubble radius $H_*^{-1}$ at that time: $\lambda_* = \epsilon_*\,H_*^{-1}$ with $\epsilon_* \leq 1$. This wavelength is then stretched by the expansion of the universe from the time of production until now. For GW produced during the radiation era when the plasma temperature is $T_*$, and assuming a standard adiabatic thermal history for the evolution of the universe after GW production, the characteristic GW frequency today can then be written as
\be
\label{Cosmokstarg}
f_c \simeq \frac{1.6 \times 10^{-4} \, \mathrm{Hz}}{\epsilon_*} \, \left(\frac{T_*}{1 \, \mathrm{TeV}}\right) \, 
\left(\frac{g_*}{100}\right)^{1/6}
\ee
where $g_*$ is the number of relativistic degrees of freedom at the temperature $T_*$. The parameter $\epsilon_* \leq 1$ depends on the dynamics of the particular GW source under consideration. For a first-order phase transition, one may have for instance $\epsilon_* \sim 10^{-3} - 1$. In that case, Eq.~(\ref{Cosmokstarg}) shows that GW produced around the electroweak scale may fall right into the mHz frequency band of eLISA. In the case of cosmic strings, GW are produced continually during the cosmological evolution (over a wide range of values of $T_*$), so that the present-day spectrum covers a very wide range of frequencies. We will discuss these two sources in Sections~\ref{CosmoPT} and \ref{CosmoStrings}. 

Several other cosmological sources of GW are reviewed in detail in \cite{Binetruy:2012ze}. One source that is very well motivated by other observations is inflation. During inflation, quantum fluctuations of the graviton field are parametrically amplified~(Ref.~\cite{Grishchuk:1974ny}) into tensor perturbations at super-Hubble scales by the quasi-exponential expansion of the universe~(Ref.~\cite{Starobinsky:1979ty}). These tensor perturbations become standard GW when they re-enter the Hubble radius in the course of the post-inflationary evolution, leading to a very broad spectrum today. If inflation occurs at sufficiently high energy scales (not much below $E_{infl} \sim 10^{16}$ GeV), the resulting GW in the frequency range $f \sim 10^{-18} - 10^{-16}\,\mathrm{Hz}$ may be indirectly detected in the future through their effect on the CMB polarization~(Refs.~\cite{Kamionkowski:1996zd}, \cite{Seljak:1996gy}). In the vast majority of cases, the current CMB bound on the inflationary GW leads already to the upper bound $h^2 \Omega_{\mathrm{gw}} \lsim 10^{-15}$ on their amplitude at frequencies accessible by direct detection experiments, see e.g.~\cite{Smith:2005mm}. This is much below the reach of eLISA and ground-based experiments. 

Under certain circumstances, however, the amplitude of the inflationary GW background may be enhanced in the direct detection frequency range, while satisfying the CMB constraint at low frequencies. This occurs in some inflationary models where particles are abundantly produced during inflation, see~\cite{Cook:2011hg}, \cite{Barnaby:2011qe}. Another possibility is that the early universe was dominated during some epoch after inflation by a fluid stiffer than radiation, i.e. with an equation of state $w > 1/3$~(see \cite{Buonanno:2003th} and references therein). A GW background with an enhanced amplitude at high frequencies may also be produced in some alternatives to inflation like the pre-big-bang scenario~(see \cite{Buonanno:2003th} and references therein). The GW produced by these processes may be observable by direct detection experiments, preferentially ground-based experiments since they probe higher frequencies. 

Another privileged epoch for GW production is the end of inflation, when the potential energy density driving inflation is converted into the thermal bath of the Hot Big Bang in the course of reheating. In many inflationary models, reheating starts with an explosive and non-perturbative decay of the inflaton condensate into very large fluctuations of itself and other bosonic fields coupled to it - a process called preheating. The large field fluctuations amplified by preheating can produce a significant amount of GW~(Refs.~\cite{Khlebnikov:1997di}, \cite{Easther:2006gt}, \cite{Easther:2006vd}, \cite{GarciaBellido:2007dg}, \cite{GarciaBellido:2007af}, \cite{Dufaux:2007pt}, \cite{Dufaux:2008dn}, \cite{Dufaux:2010cf}, \cite{Brax:2010ai}). The peak frequency of these GW today depends on the energy scale of inflation: it may fall into the frequency range of ground-based interferometers if $E_{infl} \lsim 10^{11}$ GeV and in the eLISA frequency range if $E_{infl} \lsim 10^{7}$ GeV. Thus the GW produced by preheating may be observable if inflation occurs at a sufficiently small energy scale, which is complementary to the GW produced by inflation itself. GW can also be produced by the non-perturbative decay of scalar field condensates other than the inflaton, in particular in supersymmetric theories~(Ref.~\cite{Dufaux:2009wn}). In this case, the peak GW frequency today is not directly related to the energy scale of inflation but rather to the scale of supersymmetry breaking.

Several potential sources of GW are related to the spontaneous breaking of symmetries during early universe phase transitions. Besides first-order phase transitions and cosmic strings, another possibility is provided by unstable domain 
walls~(see e.g.~\cite{Gleiser:1998na}, \cite{Takahashi:2008mu}, \cite{Hiramatsu:2010yz}). These could produce GW potentially observable by eLISA, depending on the details of the underlying high-energy physics model. GW can also be produced by the self-ordering of a scalar field after a global phase transition~(see e.g.~\cite{Krauss:1991qu}, \cite{Fenu:2009qf}, \cite{Krauss:2010df}). An observable amount of GW for eLISA would require a scalar field with a very large expectation value. 

As another example, GW may also be produced by large density perturbations leading to the formation of primordial black holes (PBH), see \cite{Saito:2009jt} and references therein. The resulting spectrum is peaked at a characteristic frequency that depends on the PBH mass. For instance, eLISA could probe the formation of PBH in the mass range $10^{22} - 10^{25}$ g, which could be dark matter candidates.

In general, different cosmological sources lead to GW backgrounds with different frequency dependences, which may allow to distinguish them from each other and from astrophysical and instrumental backgrounds. It is thus important to determine the frequency dependence of the signal as accurately as possible.

\section{First-Order Phase Transitions}
\label{CosmoPT}

In the course of its adiabatic expansion, the universe may have undergone several phase transitions driven by the temperature decrease. They occur because the effective potential of scalar fields (e.g. the Higgs field) depends on their interactions with other particles in the thermal plasma. The nature of the phase transition depends on the underlying high-energy physics model. If it is first-order, then at a certain critical temperature the scalar field becomes trapped in a false vacuum state separated from the true vacuum by a potential barrier. In that case the phase transition proceeds by tunneling across the potential barrier, leading to the nucleation of true vacuum bubbles which quickly grow into the false vacuum phase and collide. This is a violent and highly inhomogeneous process that can emit a significant amount of GW, as first discussed in \cite{Witten:1984rs}, \cite{hogan}, \cite{Turner:1990rc}. 

For phase transitions that occur at the electroweak scale, we saw below Eq.~(\ref{Cosmokstarg}) that these GW may fall right into the eLISA frequency range. We will therefore focus on these in the following. In the Standard Model of particle physics, the electroweak phase transition is not first-order and it is not expected to lead to any significant production of GW. However, it can be first-order in well-motivated extensions of the Standard Model, in particular those addressing the hierarchy problem~(see \cite{Binetruy:2012ze} for references to specific models). An observable GW signal from an electroweak phase transition would thus provide a clear indication of physics beyond the Standard Model. 

GW from first-order phase transitions are produced by two main processes: bubble collision (Refs.~\cite{Kosowsky:1991ua}, \cite{Kosowsky:1992vn}, \cite{Kamionkowski:1993fg}, \cite{Caprini:2007xq}, \cite{Huber:2008hg}, \cite{Caprini:2009fx}) and magneto-hydrodynamic turbulence (Refs.~\cite{Kamionkowski:1993fg}, \cite{Kosowsky:2001xp}, \cite{Dolgov:2002ra}, \cite{Caprini:2006jb}, \cite{Caprini:2009yp}). The calculation of the resulting GW background is reviewed in detail in \cite{Binetruy:2012ze}. It depends on five main parameters: the temperature of the phase transition $T_*$, its duration $\beta^{-1}$ ($\beta$ is related to the rate of bubble nucleation and should be larger than the Hubble rate, $\beta > H_*$), its strength $\alpha = \rho_{vac} / \rho_{rad}$ ($\rho_{vac}$ is the vacuum energy density and $\rho_{rad}$ the radiation energy density at the time of the phase transition), the velocity of the bubble walls $v_b$, and an efficiency factor $\kappa$ that measures the fraction of vacuum energy that is converted into gradient energy of the scalar field and bulk kinetic energy of the fluid (as opposed to thermal energy). These five parameters are not independent of each other and must be computed in the context of a specific high-energy physics model: $T_*$, $\beta$ and $\alpha$ depend on the scalar field potential at finite temperature, while $v_b$ and $\kappa$ depend on the propagation of the bubble wall in the surrounding plasma. 

In \cite{Binetruy:2012ze}, we calculated the GW background from first-order phase transitions by combining consistently the latest results available for GW production by bubble collision (Ref.~\cite{Huber:2008hg}) and turbulence (Ref.~\cite{Caprini:2009yp}). We also used the model of bubble propagation recently developed by \cite{Espinosa:2010hh} to evaluate the parameters $v_b$ and $\kappa$ (these parameters were previously evaluated under the particular assumption of Jouguet detonation, which is not justified in most cases). This causes the GW signal to depend on the parameter $\eta$ introduced in \cite{Espinosa:2010hh}, which characterizes the friction exerted on the bubble walls by the surrounding plasma.

\begin{figure}[htb]
\begin{center}
\includegraphics[width=9cm]{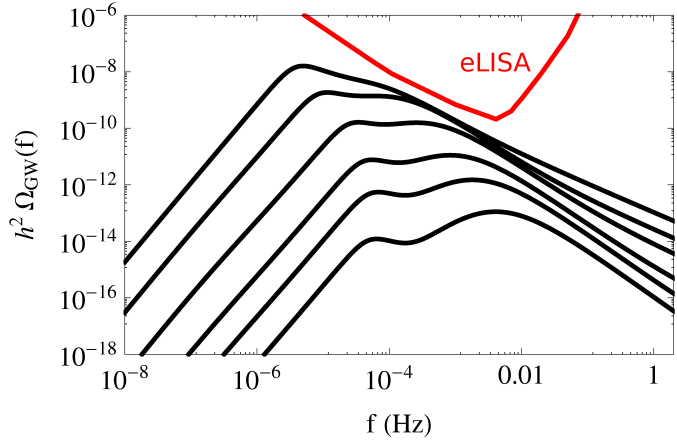}
\vspace*{-0.5cm}
\caption{\label{GWPT1_Dufaux} GW spectra from the electroweak phase transition in a model where the Higgs potential includes a dimension six operator~(Ref.~\cite{Huber:2007vva}), compared to the eLISA sensitivity curve. The GW spectra correspond to different values of the model parameters with $\alpha$ varying from $0.128$ (lower spectrum) to $2.268$ (upper spectrum) and $\eta = 1$.}
\end{center}
\end{figure}

\begin{figure}[htb]
\begin{center}
\includegraphics[width=9cm]{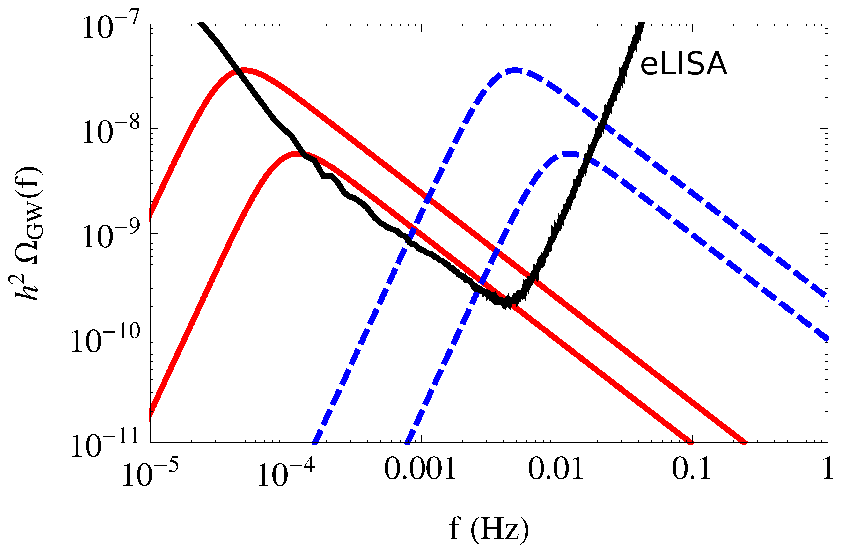}
\vspace*{-0.3cm}
\caption{\label{GWPT2_Dufaux} GW spectra from the radion phase transition in the Randall-Sundrum model~(Refs.~\cite{Randall:2006py}, \cite{Konstandin:2010cd}), compared to the eLISA sensitivity curve. The temperature of the phase transition is $T_* = 100$ GeV for the left/plain spectra and $T_* = 10^4$ GeV for the right/dashed spectra. In both cases, $\beta/H_* = 6$ for the upper spectrum and $15$ for the lower one.}
\end{center}
\end{figure}

We show the resulting GW backgrounds in two specific high-energy models with a first-order phase transition in Figs.~\ref{GWPT1_Dufaux} and \ref{GWPT2_Dufaux} (see \cite{Binetruy:2012ze} for a model-independent discussion of the detection prospects). The frequency dependence of the signal is well determined in each case. It varies as $\Omega_{gw} \propto f^3$ at low frequencies, while the high-frequency tail is given by the combination of $\Omega_{gw} \propto f^{-1}$ from bubble collision and $\Omega_{gw} \propto f^{-5/3}$ from turbulence. 

In Fig.~\ref{GWPT1_Dufaux} we show the GW spectra for different values of the parameters in a model studied in \cite{Huber:2007vva} where the Higgs potential includes a dimension six operator. For small values of $\alpha$ (lower spectra), one clearly distinguishes two different peaks, a low-frequency one due to bubble collision and a high-frequency one due to turbulence. The relative contribution of these two sources depends on $\alpha$ and $\eta$. The GW amplitude increases for values of the model parameters leading to larger values of $\alpha$, but the peak of the spectrum is then also shifted towards smaller frequencies. Furthermore, the phase transition cannot be arbitrarily strong in this model (otherwise the transition never completes). As a consequence, the amplitude of the GW signal saturates below the eLISA noise curve. 

More promising is the phase transition of the radion field in the Randall-Sundrum model, which has been studied in \cite{Randall:2006py} and \cite{Konstandin:2010cd}. In that case, the phase transition is very strong, $\alpha \gg 1$. The GW signal is then dominated by bubble collision with $\kappa \simeq 1$ and $v_b \simeq 1$. Furthermore, the amplitude of the GW background becomes independent of $\alpha$ for $\alpha \gg 1$. The signal then depends only on two parameters: $\beta / H_*$ and $T_*$, which can take a wide range of values in this model. In Fig.~\ref{GWPT2_Dufaux}, we show the GW spectra for 
$T_* = 100$ GeV (left spectra) and $T_* = 10^4$ GeV (right spectra). In both cases, the GW amplitude increases with the duration of the phase transition $\beta^{-1}$. A large signal-to-noise ratio for eLISA is obtained for slow enough phase transitions with $T_* \gsim 10^3$ GeV. For such large values of $T_*$, the model is beyond the reach of the LHC, so eLISA might be the only way to probe it.

\section{Cosmic Strings}
\label{CosmoStrings}

Cosmic strings (Ref.~\cite{VS}) are linear concentrations of energy with a cosmological size that arise in a variety of high-energy physics models, both in field theory (grand unification, supersymmetry, ...) and in string theory (Ref.~\cite{Polchinski:2004ia}). They are relics from phase transitions occurring at the end of inflation or during the thermal evolution of the universe. Once produced, a network of stable cosmic strings evolves towards a self-similar regime characterized by a continuous energy loss: when long string segments cross each other and reconnect, they form smaller cosmic string loops, which oscillate relativistically and decay away by emitting gravitational waves. The net result is that a significant fraction of the cosmic string energy density is continually converted into GW via the production of loops, from the very early universe until the present epoch. The resulting GW background (Refs.~\cite{Vilenkin:1981bx}, \cite{Hogan:1984is}, \cite{Vachaspati:1984gt}, \cite{Caldwell:1991jj}, \cite{Siemens:2006yp}, \cite{DePies:2007bm}) covers a very wide frequency range and can be looked for by different experiments, in particular pulsar timing arrays, ground-based interferometers and eLISA. In addition to this background, the GW signal from cosmic strings includes also strong and infrequent bursts produced by specific loop configurations called cusps and kinks (Refs.~\cite{Damour:2001bk}, \cite{Damour:2004kw}). These GW bursts can be looked for individually and should not be included in the computation of the stationary and nearly Gaussian background.   

\begin{figure}[htb]
\begin{center}
\includegraphics[width=15cm]{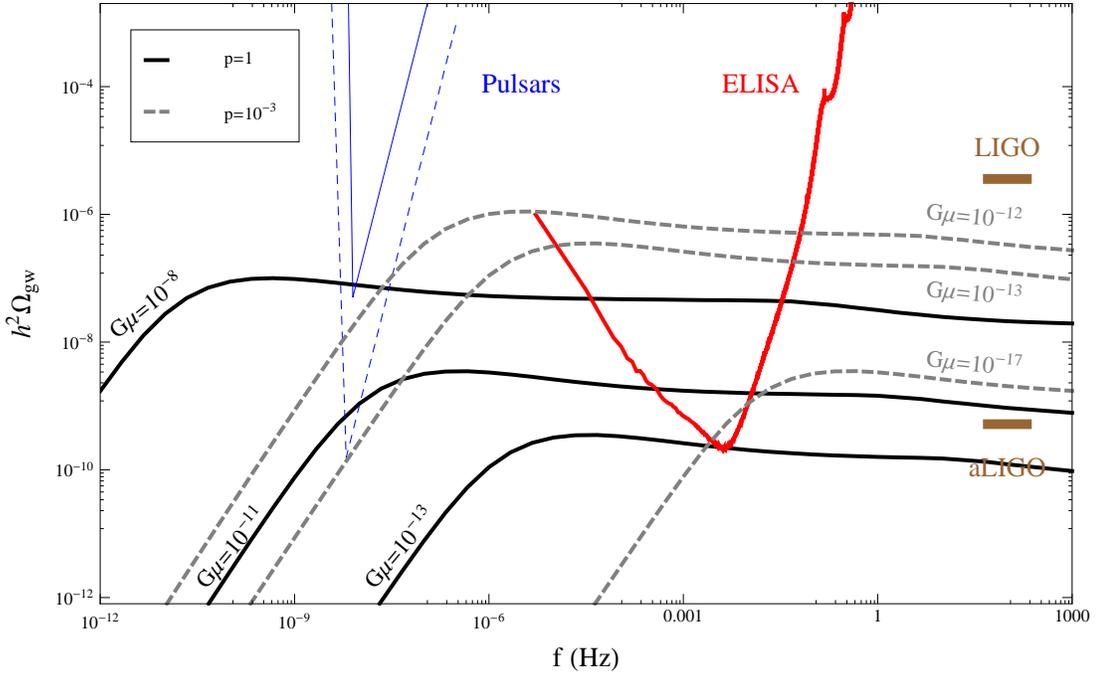}
\vspace*{-0.6cm}
\caption{\label{Strings1_Dufaux} GW background from cosmic strings compared to observational sensitivities, for different values of the cosmic string parameters in the case of long-lived loops (see the main text for details).}
\end{center}
\end{figure}

\begin{figure}[htb]
\begin{center}
\includegraphics[width=12cm]{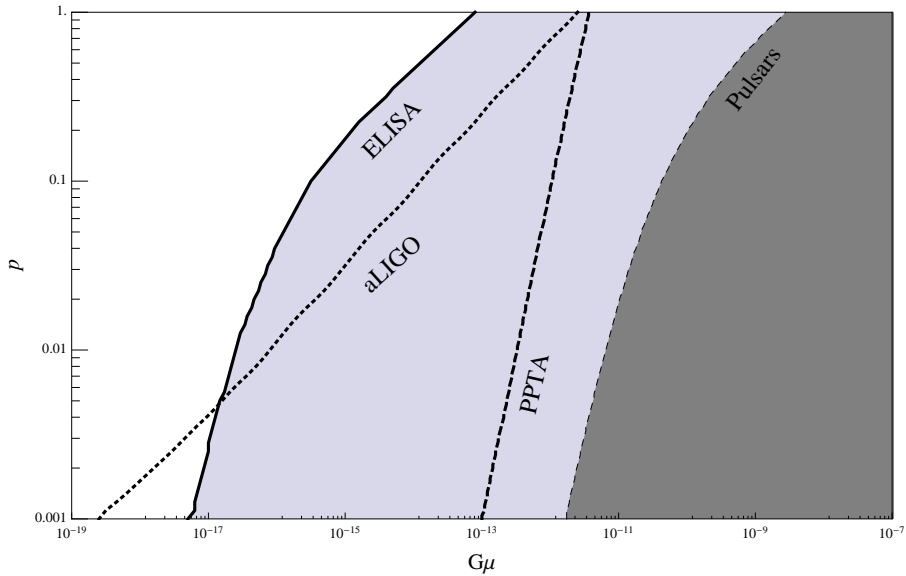}
\vspace*{-0.3cm}
\caption{\label{Strings2_Dufaux} Regions of the parameter space of cosmic strings that can be probed by different observations (at the right of each curve), in the case of long-lived loops (see the main text for details).}
\end{center}
\end{figure}

The GW signal from cosmic strings varies with two fundamental parameters that depend on the underlying high-energy physics model: the string tension $\mu$ (which is equal to their energy per unit length in the simplest models that we consider below) and the probability $p$ that two string segments reconnect when they cross each other. The dimensionless parameter $G \mu$ (where $G$ is Newton constant) must be sufficiently small to satisfy observational constraints on the gravitational effects of cosmic strings (CMB, gravitational lensing, ...)~\footnote{The production of particles by cosmic strings may also lead to complementary constraints on the string tension, see e.g.~\cite{Dufaux:2011da}, \cite{Dufaux:2012np}.}. The reconnection probability is usually $p = 1$ for cosmic strings in field theories, but it can be  smaller for cosmic strings in string theory, which increases the string energy density $\rho_s$. Simple arguments suggest that $\rho_s \propto 1/p$ (see e.g.~\cite{Damour:2004kw}), but a more complicated dependence has also been found in e.g.~\cite{Avgoustidis:2005nv}. We consider for definiteness $\rho_s \propto 1/p$ below, but the results can be directly generalized to the case where 
$\rho_s \propto 1/f(p)$ with an arbitrary function $f(p)$ by simply replacing $p$ by $f(p)$ in Figs.~\ref{Strings1_Dufaux} and \ref{Strings2_Dufaux}, see \cite{Binetruy:2012ze} for more details. 

The GW background from cosmic strings depends also strongly on the typical size of the loops when they are produced, which is still uncertain by several orders of magnitude. The latest simulations~(Ref.~\cite{BlancoPillado:2011dq}) indicate that the loops are produced with a size about ten times smaller than the Hubble radius. The loops are then long-lived compared to the Hubble time. For definiteness, we will focus on this case below ; other cases are discussed in detail in \cite{Binetruy:2012ze}. Also uncertain is the typical GW spectrum emitted by each individual loops, which depends in particular on the presence of cusps and kinks. However, we showed in \cite{Binetruy:2012ze} that this uncertainty does not strongly affect the predictions for the GW background once the total power ($P_{gw} \sim 50 G \mu^2$) emitted in GW by each loop is specified. Compared to previous works, we also improved in \cite{Binetruy:2012ze} the modeling of the cosmological evolution, because it affects the frequency dependence of the GW background.

The resulting GW background for long-lived loops is shown in Fig.~\ref{Strings1_Dufaux} for different values of $G \mu$ and $p$. The frequency dependence of the signal is well determined. The high-frequency part of the spectrum (which is produced during the radiation era) is almost flat over many decades of frequency (small deviations from an exactly flat spectrum arise because of the variation of the number of relativistic species during the expansion of the universe). The spectrum slightly increases at lower frequencies and then falls off as $\Omega_{gw} \propto f^{3/2}$ at very small frequencies.

An important feature of the GW background from cosmic strings is that, because it covers a very wide frequency range, it can be observable by different experiments. We show the sensitivity of eLISA, ground-based interferometers and pulsar timing arrays in Fig.~\ref{Strings1_Dufaux}. For ground-based interferometers, we consider the upper bound set by the LIGO S5 run and the expected sensitivity of Advanced LIGO, both taken from \cite{Abbott:2009ws}. For pulsar timing experiments, we consider the upper bound obtained in \cite{Jenet:2006sv} and the future sensitivity of the complete Parks Pulsar Timing Array (PPTA) estimated in that reference. 

In Fig.~\ref{Strings2_Dufaux}, we show the regions of the parameter space that can be probed by these experiments, still in the case of long-lived loops. The darkest region is already excluded by current pulsar bounds. In a large part of the parameter space, the GW signal could be simultaneously observed by eLISA and at least one other experiment. This would certainly improve the characterization of the signal and the ability to distinguish it from other stochastic backgrounds. Other regions of the parameter space are only accessible to eLISA, which could thus open a new window on the cosmic string parameters. In particular, eLISA would be able to probe most of the parameter space for cosmic strings that arise in the simplest string theory models.

\section{Conclusion} 
\label{CosmoConclu}

We have investigated the scientific potential of the eLISA mission regarding the detection of cosmological GW backgrounds. We have considered different cosmological sources that could lead to an observable signal, with a special emphasis on first-order phase transitions and cosmic strings. Each of these two sources produces a GW background with a distinct and well determined frequency dependence, which allows to distinguish the signal from other cosmological or astrophysical backgrounds and from the instrumental noise. In both cases, the signal may lie well above the eLISA noise curve, depending on the microphysics of the source. The detection of a cosmological GW background by eLISA is of course not guaranteed: it could just constrain cosmological sources. However, such a detection is at least possible, and this would be a major discovery for our understanding of the very early universe and the underlying high-energy physics.

\emph{Acknowledgments.} It is a pleasure to thank Pierre Bin\'etruy, Alejandro Boh\'e and Chiara Caprini for collaborations and the organizers of the ninth LISA symposium for a very interesting conference.


\bibliography{MyBibli_Dufaux}

\begin{thebibliography}{}
\expandafter\ifx\csname natexlab\endcsname\relax\def\natexlab#1{#1}\fi
\expandafter\ifx\csname url\endcsname\relax
  \def\url#1{\texttt{#1}}\fi
\expandafter\ifx\csname urlprefix\endcsname\relax\def\urlprefix{URL }\fi
\providecommand{\eprint}[2][]{\url{#2}}

\bibitem[{Abbott et~al.(2009)}]{Abbott:2009ws}
Abbott, B., et~al. (LIGO Scientific Collaboration, VIRGO Collaboration) 2009,
  Nature, 460, 990. \eprint{0910.5772}

\bibitem[{Allen(1996)}]{Allen:1996vm}
Allen, B. 1996. \eprint{gr-qc/9604033}

\bibitem[{Amaro-Seoane et~al.(2012{\natexlab{a}})Amaro-Seoane, Aoudia, Babak,
  Binetruy, Berti et~al.}]{AmaroSeoane:2012km}
Amaro-Seoane, P., Aoudia, S., Babak, S., Binetruy, P., Berti, E., et~al.
  2012{\natexlab{a}}. \eprint{1201.3621}

\bibitem[{Amaro-Seoane et~al.(2012{\natexlab{b}})Amaro-Seoane, Aoudia, Babak,
  Binetruy, Berti et~al.}]{AmaroSeoane:2012je}
--- 2012{\natexlab{b}}, Class.Quant.Grav., 29, 124016. \eprint{1202.0839}

\bibitem[{Avgoustidis \& Shellard(2006)}]{Avgoustidis:2005nv}
Avgoustidis, A., \& Shellard, E. 2006, Phys.Rev., D73, 041301.
  \eprint{astro-ph/0512582}

\bibitem[{Barnaby et~al.(2012)Barnaby, Pajer, \& Peloso}]{Barnaby:2011qe}
Barnaby, N., Pajer, E., \& Peloso, M. 2012, Phys.Rev., D85, 023525.
  \eprint{1110.3327}

\bibitem[{Binetruy et~al.(2012)Binetruy, Bohe, Caprini, \&
  Dufaux}]{Binetruy:2012ze}
Binetruy, P., Bohe, A., Caprini, C., \& Dufaux, J.-F. 2012, JCAP, 1206, 027.
  \eprint{1201.0983}

\bibitem[{Blanco-Pillado et~al.(2011)Blanco-Pillado, Olum, \&
  Shlaer}]{BlancoPillado:2011dq}
Blanco-Pillado, J.~J., Olum, K.~D., \& Shlaer, B. 2011, Phys.Rev., D83, 083514.
  \eprint{1101.5173}

\bibitem[{Brax et~al.(2011)Brax, Dufaux, \& Mariadassou}]{Brax:2010ai}
Brax, P., Dufaux, J.-F., \& Mariadassou, S. 2011, Phys.Rev., D83, 103510.
  \eprint{1012.4656}

\bibitem[{Buonanno(2003)}]{Buonanno:2003th}
Buonanno, A. 2003, 855. \eprint{gr-qc/0303085}

\bibitem[{Caldwell \& Allen(1992)}]{Caldwell:1991jj}
Caldwell, R., \& Allen, B. 1992, Phys.Rev., D45, 3447

\bibitem[{Caprini \& Durrer(2006)}]{Caprini:2006jb}
Caprini, C., \& Durrer, R. 2006, Phys.Rev., D74, 063521.
  \eprint{astro-ph/0603476}

\bibitem[{Caprini et~al.(2009{\natexlab{a}})Caprini, Durrer, Konstandin, \&
  Servant}]{Caprini:2009fx}
Caprini, C., Durrer, R., Konstandin, T., \& Servant, G. 2009{\natexlab{a}},
  Phys.Rev., D79, 083519. \eprint{0901.1661}

\bibitem[{Caprini et~al.(2008)Caprini, Durrer, \& Servant}]{Caprini:2007xq}
Caprini, C., Durrer, R., \& Servant, G. 2008, Phys.Rev., D77, 124015.
  \eprint{0711.2593}

\bibitem[{Caprini et~al.(2009{\natexlab{b}})Caprini, Durrer, \&
  Servant}]{Caprini:2009yp}
--- 2009{\natexlab{b}}, JCAP, 0912, 024. \eprint{0909.0622}

\bibitem[{Cook \& Sorbo(2012)}]{Cook:2011hg}
Cook, J.~L., \& Sorbo, L. 2012, Phys.Rev., D85, 023534. \eprint{1109.0022}

\bibitem[{Damour \& Vilenkin(2001)}]{Damour:2001bk}
Damour, T., \& Vilenkin, A. 2001, Phys.Rev., D64, 064008.
  \eprint{gr-qc/0104026}

\bibitem[{Damour \& Vilenkin(2005)}]{Damour:2004kw}
--- 2005, Phys.Rev., D71, 063510. \eprint{hep-th/0410222}

\bibitem[{DePies \& Hogan(2007)}]{DePies:2007bm}
DePies, M.~R., \& Hogan, C.~J. 2007, Phys.Rev., D75, 125006.
  \eprint{astro-ph/0702335}

\bibitem[{Dolgov et~al.(2002)Dolgov, Grasso, \& Nicolis}]{Dolgov:2002ra}
Dolgov, A.~D., Grasso, D., \& Nicolis, A. 2002, Phys.Rev., D66, 103505.
  \eprint{astro-ph/0206461}

\bibitem[{Dufaux(2009)}]{Dufaux:2009wn}
Dufaux, J.-F. 2009, Phys.Rev.Lett., 103, 041301. \eprint{0902.2574}

\bibitem[{Dufaux(2012{\natexlab{a}})}]{Dufaux:2011da}
--- 2012{\natexlab{a}}, Phys.Rev.Lett., 109, 011601. \eprint{1109.5121}

\bibitem[{Dufaux(2012{\natexlab{b}})}]{Dufaux:2012np}
--- 2012{\natexlab{b}}. \eprint{1201.4850}

\bibitem[{Dufaux et~al.(2007)Dufaux, Bergman, Felder, Kofman, \&
  Uzan}]{Dufaux:2007pt}
Dufaux, J.~F., Bergman, A., Felder, G.~N., Kofman, L., \& Uzan, J.-P. 2007,
  Phys.Rev., D76, 123517. \eprint{0707.0875}

\bibitem[{Dufaux et~al.(2009)Dufaux, Felder, Kofman, \& Navros}]{Dufaux:2008dn}
Dufaux, J.-F., Felder, G., Kofman, L., \& Navros, O. 2009, JCAP, 0903, 001.
  \eprint{0812.2917}

\bibitem[{Dufaux et~al.(2010)Dufaux, Figueroa, \&
  Garcia-Bellido}]{Dufaux:2010cf}
Dufaux, J.-F., Figueroa, D.~G., \& Garcia-Bellido, J. 2010, Phys.Rev., D82,
  083518. \eprint{1006.0217}

\bibitem[{Easther et~al.(2007)Easther, Giblin, \& Lim}]{Easther:2006vd}
Easther, R., Giblin, J., John~T., \& Lim, E.~A. 2007, Phys.Rev.Lett., 99,
  221301. \eprint{astro-ph/0612294}

\bibitem[{Easther \& Lim(2006)}]{Easther:2006gt}
Easther, R., \& Lim, E.~A. 2006, JCAP, 0604, 010. \eprint{astro-ph/0601617}

\bibitem[{Espinosa et~al.(2010)Espinosa, Konstandin, No, \&
  Servant}]{Espinosa:2010hh}
Espinosa, J.~R., Konstandin, T., No, J.~M., \& Servant, G. 2010, JCAP, 1006,
  028. \eprint{1004.4187}

\bibitem[{Fenu et~al.(2009)Fenu, Figueroa, Durrer, \&
  Garcia-Bellido}]{Fenu:2009qf}
Fenu, E., Figueroa, D.~G., Durrer, R., \& Garcia-Bellido, J. 2009, JCAP, 0910,
  005. \eprint{0908.0425}

\bibitem[{Garcia-Bellido \& Figueroa(2007)}]{GarciaBellido:2007dg}
Garcia-Bellido, J., \& Figueroa, D.~G. 2007, Phys.Rev.Lett., 98, 061302.
  \eprint{astro-ph/0701014}

\bibitem[{Garcia-Bellido et~al.(2008)Garcia-Bellido, Figueroa, \&
  Sastre}]{GarciaBellido:2007af}
Garcia-Bellido, J., Figueroa, D.~G., \& Sastre, A. 2008, Phys.Rev., D77,
  043517. \eprint{0707.0839}

\bibitem[{Gleiser \& Roberts(1998)}]{Gleiser:1998na}
Gleiser, M., \& Roberts, R. 1998, Phys.Rev.Lett., 81, 5497.
  \eprint{astro-ph/9807260}

\bibitem[{Grishchuk(1975)}]{Grishchuk:1974ny}
Grishchuk, L. 1975, Sov.Phys.JETP, 40, 409

\bibitem[{Hiramatsu et~al.(2010)Hiramatsu, Kawasaki, \&
  Saikawa}]{Hiramatsu:2010yz}
Hiramatsu, T., Kawasaki, M., \& Saikawa, K. 2010, JCAP, 1005, 032.
  \eprint{1002.1555}

\bibitem[{Hogan \& Rees(1984)}]{Hogan:1984is}
Hogan, C., \& Rees, M. 1984, Nature, 311, 109

\bibitem[{Hogan(1986)}]{hogan}
Hogan, C.~J. 1986, Mon.Not.Roy.Astron.Soc., 218

\bibitem[{Hogan(2006)}]{Hogan:2006va}
--- 2006, AIP Conf.Proc., 873, 30. \eprint{astro-ph/0608567}

\bibitem[{Hogan \& Bender(2001)}]{Hogan:2001jn}
Hogan, C.~J., \& Bender, P.~L. 2001, Phys.Rev., D64, 062002.
  \eprint{astro-ph/0104266}

\bibitem[{Huber \& Konstandin(2008{\natexlab{a}})}]{Huber:2008hg}
Huber, S.~J., \& Konstandin, T. 2008{\natexlab{a}}, JCAP, 0809, 022.
  \eprint{0806.1828}

\bibitem[{Huber \& Konstandin(2008{\natexlab{b}})}]{Huber:2007vva}
--- 2008{\natexlab{b}}, JCAP, 0805, 017. \eprint{0709.2091}

\bibitem[{Jenet et~al.(2006)Jenet, Hobbs, van Straten, Manchester, Bailes
  et~al.}]{Jenet:2006sv}
Jenet, F.~A., Hobbs, G., van Straten, W., Manchester, R., Bailes, M., et~al.
  2006, Astrophys.J., 653, 1571. \eprint{astro-ph/0609013}

\bibitem[{Kamionkowski et~al.(1997)Kamionkowski, Kosowsky, \&
  Stebbins}]{Kamionkowski:1996zd}
Kamionkowski, M., Kosowsky, A., \& Stebbins, A. 1997, Phys.Rev.Lett., 78, 2058.
  \eprint{astro-ph/9609132}

\bibitem[{Kamionkowski et~al.(1994)Kamionkowski, Kosowsky, \&
  Turner}]{Kamionkowski:1993fg}
Kamionkowski, M., Kosowsky, A., \& Turner, M.~S. 1994, Phys.Rev., D49, 2837.
  \eprint{astro-ph/9310044}

\bibitem[{Khlebnikov \& Tkachev(1997)}]{Khlebnikov:1997di}
Khlebnikov, S., \& Tkachev, I. 1997, Phys.Rev., D56, 653.
  \eprint{hep-ph/9701423}

\bibitem[{Konstandin et~al.(2010)Konstandin, Nardini, \&
  Quiros}]{Konstandin:2010cd}
Konstandin, T., Nardini, G., \& Quiros, M. 2010, Phys.Rev., D82, 083513.
  \eprint{1007.1468}

\bibitem[{Kosowsky et~al.(2002)Kosowsky, Mack, \&
  Kahniashvili}]{Kosowsky:2001xp}
Kosowsky, A., Mack, A., \& Kahniashvili, T. 2002, Phys.Rev., D66, 024030.
  \eprint{astro-ph/0111483}

\bibitem[{Kosowsky \& Turner(1993)}]{Kosowsky:1992vn}
Kosowsky, A., \& Turner, M.~S. 1993, Phys.Rev., D47, 4372.
  \eprint{astro-ph/9211004}

\bibitem[{Kosowsky et~al.(1992)Kosowsky, Turner, \& Watkins}]{Kosowsky:1991ua}
Kosowsky, A., Turner, M.~S., \& Watkins, R. 1992, Phys.Rev., D45, 4514

\bibitem[{Krauss(1992)}]{Krauss:1991qu}
Krauss, L.~M. 1992, Phys.Lett., B284, 229

\bibitem[{Krauss et~al.(2010)Krauss, Jones-Smith, Mathur, \&
  Dent}]{Krauss:2010df}
Krauss, L.~M., Jones-Smith, K., Mathur, H., \& Dent, J. 2010, Phys.Rev., D82,
  044001. \eprint{1003.1735}

\bibitem[{Maggiore(2000)}]{Maggiore:1999vm}
Maggiore, M. 2000, Phys.Rept., 331, 283. \eprint{gr-qc/9909001}

\bibitem[{Polchinski(2004)}]{Polchinski:2004ia}
Polchinski, J. 2004, 229. \eprint{hep-th/0412244}

\bibitem[{Randall \& Servant(2007)}]{Randall:2006py}
Randall, L., \& Servant, G. 2007, JHEP, 0705, 054. \eprint{hep-ph/0607158}

\bibitem[{Saito \& Yokoyama(2010)}]{Saito:2009jt}
Saito, R., \& Yokoyama, J. 2010, Prog.Theor.Phys., 123, 867. \eprint{0912.5317}

\bibitem[{Seljak \& Zaldarriaga(1997)}]{Seljak:1996gy}
Seljak, U., \& Zaldarriaga, M. 1997, Phys.Rev.Lett., 78, 2054.
  \eprint{astro-ph/9609169}

\bibitem[{Siemens et~al.(2007)Siemens, Mandic, \& Creighton}]{Siemens:2006yp}
Siemens, X., Mandic, V., \& Creighton, J. 2007, Phys.Rev.Lett., 98, 111101.
  \eprint{astro-ph/0610920}

\bibitem[{Smith et~al.(2006{\natexlab{a}})Smith, Kamionkowski, \&
  Cooray}]{Smith:2005mm}
Smith, T.~L., Kamionkowski, M., \& Cooray, A. 2006{\natexlab{a}}, Phys.Rev.,
  D73, 023504. \eprint{astro-ph/0506422}

\bibitem[{Smith et~al.(2006{\natexlab{b}})Smith, Pierpaoli, \&
  Kamionkowski}]{Smith:2006nka}
Smith, T.~L., Pierpaoli, E., \& Kamionkowski, M. 2006{\natexlab{b}},
  Phys.Rev.Lett., 97, 021301. \eprint{astro-ph/0603144}

\bibitem[{Starobinsky(1979)}]{Starobinsky:1979ty}
Starobinsky, A.~A. 1979, JETP Lett., 30, 682

\bibitem[{Takahashi et~al.(2008)Takahashi, Yanagida, \&
  Yonekura}]{Takahashi:2008mu}
Takahashi, F., Yanagida, T., \& Yonekura, K. 2008, Phys.Lett., B664, 194.
  \eprint{0802.4335}

\bibitem[{Tinto et~al.(2001)Tinto, Armstrong, \& Estabrook}]{Tinto:2001ii}
Tinto, M., Armstrong, J., \& Estabrook, F. 2001, Phys.Rev., D63, 021101

\bibitem[{Turner \& Wilczek(1990)}]{Turner:1990rc}
Turner, M.~S., \& Wilczek, F. 1990, Phys.Rev.Lett., 65, 3080

\bibitem[{Vachaspati \& Vilenkin(1985)}]{Vachaspati:1984gt}
Vachaspati, T., \& Vilenkin, A. 1985, Phys.Rev., D31, 3052

\bibitem[{van Haasteren et~al.(2011)van Haasteren, Levin, Janssen, Lazaridis,
  Stappers et~al.}]{vanHaasteren:2011ni}
van Haasteren, R., Levin, Y., Janssen, G., Lazaridis, K., Stappers, M. K.~B.,
  et~al. 2011. \eprint{1103.0576}

\bibitem[{Vilenkin(1981)}]{Vilenkin:1981bx}
Vilenkin, A. 1981, Phys.Lett., B107, 47

\bibitem[{Vilenkin \& Shellard(1994)}]{VS}
Vilenkin, A., \& Shellard, E. 1994, {\emph{Cosmic Strings and Other Topological
  Defects} (Cambridge University Press)}

\bibitem[{Witten(1984)}]{Witten:1984rs}
Witten, E. 1984, Phys.Rev., D30, 272

\end{thebibliography}
\bibliographystyle{asp2010}

\end{document}